\documentclass[a4paper,conference]{IEEEtran}
\ifCLASSINFOpdf
\else
\fi
\usepackage{graphicx}
\usepackage{amsmath}
\usepackage{subfig}
\usepackage{url}

\begin{document}

\title{Intensity-Sensitive Similarity Indexes for Image Quality Assessment}

\author{\IEEEauthorblockN{Xiaotong Li}
\IEEEauthorblockA{Oxford e-Research Centre\\
Department of Engineering Science\\
University of Oxford\\
Oxford, UK\\
Email: xiaotong.li@eng.ox.ac.uk}
\and
\IEEEauthorblockN{Wesley Armour}
\IEEEauthorblockA{Oxford e-Research Centre\\
Department of Engineering Science\\
University of Oxford\\
Oxford, UK\\
Email: wes.armour@oerc.ox.ac.uk}
}

\maketitle

\footnote{\copyright 2022 IEEE. Personal use of this material is permitted. Permission from IEEE must be obtained for all other uses, in any current or future media, including reprinting/republishing this material for advertising or promotional purposes, creating new collective works, for resale or redistribution to servers or lists, or reuse of any copyrighted component of this work in other works.

X. Li and W. Armour, "Intensity-Sensitive Similarity Indexes for Image Quality Assessment," 2022 26th International Conference on Pattern Recognition (ICPR), 2022, pp. 1975-1981, doi: 10.1109/ICPR56361.2022.9956093.

https://ieeexplore.ieee.org/document/9956093}

\begin{abstract}
The importance of Image quality assessment (IQA) is ever increasing due to the fast paced advances in imaging technology and computer vision. Among the  numerous IQA methods, Structural SIMilarity (SSIM) index and its variants are better matched to the perceived quality of the human visual system. However, SSIM methods are insufficiently sensitive, when images contain low information, where the important information only occupies a low proportion of the image while most of the image is noise-like, which is common in scientific data. Therefore, we propose two new IQA methods, InTensity Weighted SSIM index and Low-Information Similarity Index, for such low information images. In addition, auxiliary indexes are proposed to assist with the assessment. The application of these new IQA methods to natural images and field-specific images, such as radio astronomical images, medical images, and remote sensing images, are also demonstrated. The results show that our IQA methods perform better than state-of-the-art SSIM methods for differences in high-intensity parts of the input images and have similar performance to that of the original and gradient-based SSIM for differences in low-intensity parts. Different similarity indexes are suitable for different applications, which we demonstrate in our results.
\end{abstract}

\IEEEpeerreviewmaketitle

\section{Introduction}
With the fast paced development of imaging technology and computer vision in recent decades, there is an increased demand for image quality assessment (IQA). Comparing with subjective IQA which is time-consuming and expensive, objective IQA \cite{IQA1} is in favour with scholars especially in the era of big data. The most widely used full-reference (FR) IQA metrics are mean squared error (MSE), peak signal-to-noise ratio (PSNR), and structural similarity (SSIM) \cite{SSIM1,SSIM12,SSIM13}. Herein, we focus on discussing SSIM and its variants, since SSIMs are better matched to the perceived quality of the human visual system (HVS) comparing with other FR-IQA metrics; and to assess the image quality of scientific data, the scalar values of PSNR and MSE will reduce information.

The original SSIM is expressed as Equation (\ref{SSIMa}), where \textbf{\textit{x}} and \textbf{\textit{y}} are two input images, $\mu_{x}$ ($\mu_{y}$) indicates the mean intensity of \textbf{\textit{x}} (\textbf{\textit{y}}), $\sigma_{x}$ ($\sigma_{y}$) indicates the standard deviation of \textbf{\textit{x}} (\textbf{\textit{y}}), and $\sigma_{xy}$ indicates the cross covariance of the inputs.

\begin{equation}
\label{SSIMa}
    SSIM\left( {x,y} \right) = \frac{\left( {2\mu_{x}\mu_{y} + C_{1}} \right)\left( {2\sigma_{xy} + C_{2}} \right)}{\left( {\mu_{x}^{2} + \mu_{y}^{2} + C_{1}} \right)\left( {\sigma_{x}^{2} + \sigma_{y}^{2} + C_{2}} \right)}
\end{equation}

Since the appearance of SSIM, a series of variants have been devised. To evaluate similarity at different resolutions or scales, multi-scale structural similarity (MS-SSIM) \cite{SSIM4} has generalised the single-scale SSIM to multi-scale cases. To better match the HVS, visual attention information has been added to the IQA metrics \cite{SSIM11}.

\begin{table*}[!t]
\begin{center}
\caption{Proportion relationship between original and weighted versions of SSIM.}
\begin{tabular}{|c||c||c||c||c|} 
 \hline
SSIM & Total amount of value & Weight of each pixel & Amount of value of each pixel & Verification \\
 \hline
Original & $N$ & $\frac{1}{N}$ & $1$ & ${\sum\limits_{i = 1}^{N}1} = N$\\
\hline
Weighted & $N$ & $f\left( x_{i} \right)$ & $f\left( x_{i} \right)N$ & ${\sum\limits_{i = 1}^{N}{f\left( x_{i} \right)N}} = N{\sum\limits_{i = 1}^{N}{f\left( x_{i} \right)}} = N$\\
\hline
\end{tabular}
\end{center}
\end{table*}

Some variants of SSIM attach different importance to different information. To deal with blurred images, a gradient-based structural similarity (G-SSIM) \cite{SSIM5} has been developed, which considers the edge of an image as the most important structure information. Following that, several improvements have been made \cite{SSIM6,SSIM7} to increase its performance on blurred and noisy images. Additive and spatial pooling methods have also been used to further improve SSIM and G-SSIM \cite{SSIM9}. In addition, by weighting image quality with visual importance, fixation-SSIM (F-SSIM) and percentile-SSIM (P-SSIM) as well as their combination, PF-SSIM, have been proposed \cite{SSIM8}. Moreover, information content weighted SSIM (IW-SSIM) \cite{SSIM10} has been devised based on a Gaussian scale mixture \cite{GSM} model. Besides, complex wavelet structural similarity (CW-SSIM) \cite{SSIM3,SSIM2} generalised the SSIM to the complex wavelet domain, where the non-structural geometric distortions are considered to be less important.

\begin{figure}[!t]
    \centering
    \includegraphics[width=2.5in]{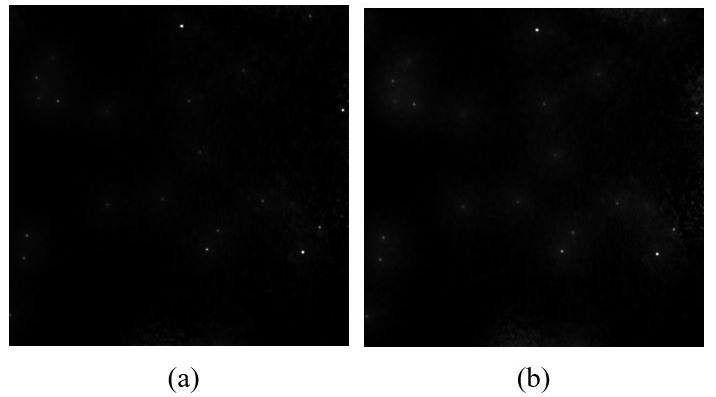}
\caption[Restored GLEAM.]{Restored images (logarithmic) of GLEAM at (a) 2\% and (b) 5\% noise levels. The differences between them are more visible in logarithmic than in linear representations.}
\end{figure}

Currently, most SSIM methods aim at assessing images from the aspect of HVS. However, in scientific applications, when the images only contain low information, SSIM methods are no longer sensitive. Take the assessment of radio astronomical (RA) images (Fig. 1) to illustrate the problem. In RA images, effective information is contained in the celestial sources. Since sources only occupy a few parts of the image while most parts of the image are noise-like, most RA images are low-information compared to natural images. It is significant to assess the similarity between the informative parts.

In the example, data in a simulated dataset, called GLEAM \cite{GLEAM1,GLEAM2} is used. The restored sky brightness images of GLEAM at different noise levels are shown in Fig. 1. Herein, the noise level is defined as a percentage of the intensity of the brightest source. A particular pixel is considered as noise if its intensity is smaller than the intensity corresponding to the noise level. To assess the quality between the restored images at different noise levels, the images are compared by SSIM. Because of the resemblance between the extensive noise-like parts in the images, the original SSIM of them is 0.9707, which is approaching 1, indicating that they are very similar. However, the extent of differences between the sources in the two images is required to be identified.

To address this problem, two new intensity-sensitive similarity indexes are proposed. One is InTensity Weighted (ITW) SSIM index, which is developed based on the SSIM structure. Weighting factors are added into the statistical variables to attach higher importance to high-intensity pixels. The other is Low-Information Similarity Index (LISI), which is different from the SSIM structure. It is developed based on the principle that an effective intensity-sensitive similarity index should reflect the amount of information contained in the images and be larger when the values of pixels in the two input images at the same position are more similar and both with high intensities. It is worth noting that our IQA methods are aimed at scientific data, not necessarily HVS.

\section{InTensity Weighted SSIM Index}

To assess the quality of low-information images, an effective IQA method should rely on the intensity of the pixel instead of whether it belongs to an important object. In fact, if the assessment was related to the judgement of an object, an empirical threshold would be required to distinguish objects and noise, which would introduce error at the spread edge of the objects.

\begin{figure}[!t]
    \centering
    \includegraphics[width=2.1in]{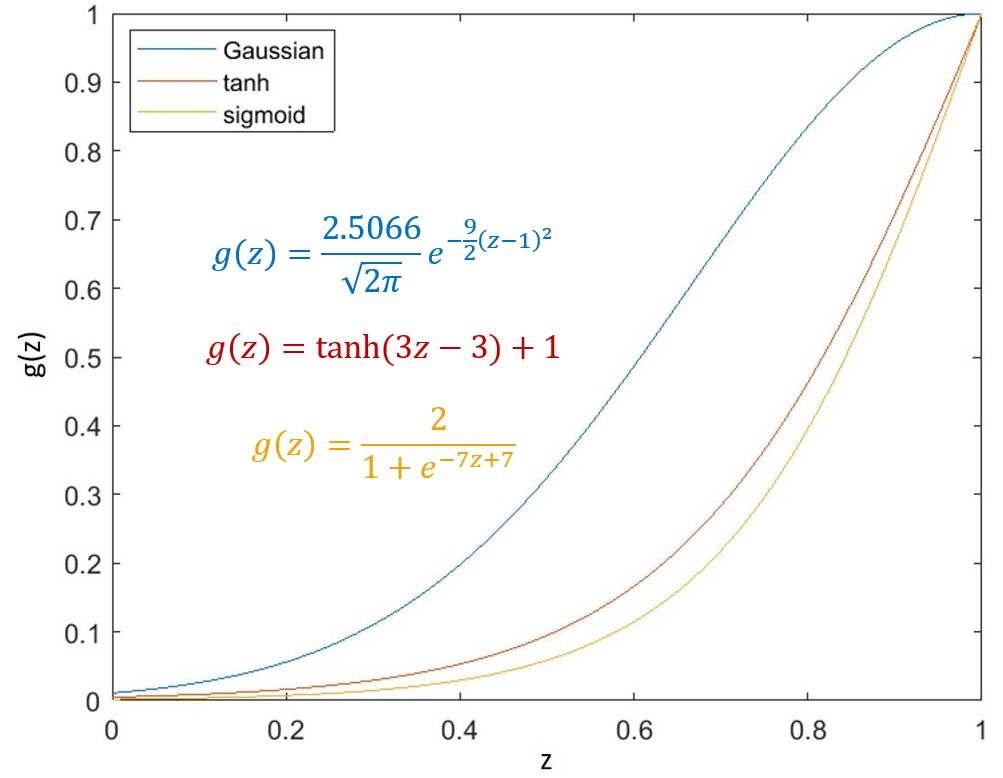}
\caption{Weighting functions.}
\end{figure}

Based on SSIM, an InTensity Weighted SSIM (ITW-SSIM) index is developed, in which weighting functions are applied to weight the importance of each pixel according to its intensity. The weighting factor is supposed to be larger for high intensity pixels and smaller for low intensity pixels. Gaussian-, tanh-, and sigmoid-based weighting functions $g\left( z \right)$ have been derived as visualised in Fig. 2, where the parameters are set so that $g\left( z \right) \in \left( 0,1 \right)$ when $z \in \left( 0,1 \right)$.

To ensure the weighting process is in accordance with the probabilistic model, the proportion relationship (Table I) between original and weighted versions should be obeyed. In Table I, $f\left( \bullet \right)$ is the weighting factor. According to probability theory, the sum of the weighting factor for each image should be 1. Thus, the relationship between weighting factor $f\left( z \right)$ and weighting function $g\left( z \right)$ is expressed as Equation (\ref{weight}).

\begin{equation}
\label{weight}
    f\left( z \right) = \frac{g\left( z \right)}{\sum\limits_{j = 1}^{N}{g\left( x_{j} \right)}}
\end{equation}

To calculate mean intensity, each pixel needs to be weighted by its importance. In the variance and the cross covariance, on the other hand, the importance is reflected in the $\mu$-term and the value of each pixel. As the $\mu$-term has already been weighted in the mean intensity, only the value of each pixel needs to be weighted. The degree of freedom should be $N-1$ in the weighted cases.

Therefore, ITW-SSIM ($ITW-SSIM \in \left\lbrack - 1,1 \right\rbrack$, which equals 1 when input images are identical) is designed as Equation (\ref{ITW}), where Equation (\ref{ITWb}) describes its variables, $\mu_{y}$ (or $\sigma_{y}$) is in similar form of $\mu_{x}$ (or $\sigma_{x}$), $N$ is the total number of pixels in the image, index $i$ indicates a pixel in the image, and the weighting factor $f\left( \bullet \right)$ is shown in Equation (\ref{weight}). The constant variables are set as $C_1=1 \times {10}^{-4}$ and $C_2=9 \times {10}^{-4}$, according to the settings in the original SSIM \cite{SSIM1}. 
\begin{subequations}
\label{ITW}
\begin{equation}
\label{ITWa}
    ITW-SSIM\left( {x,y} \right) = \frac{\left( {2\mu_{x}\mu_{y} + C_{1}} \right)\left( {2\sigma_{xy} + C_{2}} \right)}{\left( {\mu_{x}^{2} + \mu_{y}^{2} + C_{1}} \right)\left( {\sigma_{x}^{2} + \sigma_{y}^{2} + C_{2}} \right)}
\end{equation}
\begin{equation}
\label{ITWb}
\left\{ \begin{matrix}
{\mu_{x} = {\sum\limits_{i = 1}^{N}{f\left( x_{i} \right)x_{i}}}~~~~~~~~~~~~~~~~~~~~~~~~~~~~~~~~~~~~~~~~} \\
{\sigma_{x} = \left( {\frac{1}{N-1}{\sum\limits_{i = 1}^{N}\left( {f\left( x_{i} \right)Nx_{i} - \mu_{x}} \right)^{2}}} \right)^{\frac{1}{2}}~~~~~~~~~~~~~~~} \\
{\sigma_{xy} = \frac{1}{N-1}{\sum\limits_{i = 1}^{N}{\left( {f\left( x_{i} \right)Nx_{i} - \mu_{x}} \right)\left( {f\left( y_{i} \right)Ny_{i} - \mu_{y}} \right)}}} \\
\end{matrix} \right.
\end{equation}
\end{subequations}

To make the performance of ITW-SSIM independent of the range of input values, a normalisation is needed in the pre-processing. For two input images \textbf{\textit{x}} and \textbf{\textit{y}} with the same size, the normalisation should not change the relationship between the two images. Thereby, the two images should be normalised together, as shown in Equation (\ref{normalise}), where $i = 1,2,\ldots,N$.

\begin{equation}
    \label{normalise}
    \left\{ \begin{matrix}
{x_{i} = \frac{x_{i} - \min\big( \min\left( \mathbf{x} \right),\min\left( \mathbf{y} \right) \big)}{\max\big( \max\left( \mathbf{x} \right),\max\left( \mathbf{y} \right) \big) - \min\big( \min\left( \mathbf{x} \right),\min\left( \mathbf{y} \right) \big)}} \\
{y_{i} = \frac{y_{i} - \min\big( \min\left( \mathbf{x} \right),\min\left( \mathbf{y} \right) \big)}{\max\big( \max\left( \mathbf{x} \right),\max\left( \mathbf{y} \right) \big) - \min\big( \min\left( \mathbf{x} \right),\min\left( \mathbf{y} \right) \big)}} \\
\end{matrix} \right.
\end{equation}

By normalising the input images, each value in the inputs of ITW-SSIM will be between 0 and 1.

\section{Low-Information Similarity Index}

In this section we propose a Low-Information Similarity Index (LISI), which further enhances the sensitivity of the similarity index to differences in low-information images.

To assess low-information images, an effective similarity index is supposed to be larger when the values of pixels in the two input images at the same position are more similar and both with high intensities (i.e., likely belong to important objects); otherwise, the similarity index should be smaller (because there is no effective information in two pixels both representing noise even if they are similar, unless for someone who studies the noise in the image). Moreover, it is more reasonable to measure the similarity relatively, rather than absolutely. In other words, for a certain number of similar object-like pixels, the index should be different when the total number of important objects are different.

The design of LISI starts from a binary case whose schematic diagram is shown in Fig. 3. In the binary case, for each pixel, there are only two states, important or not. Based on the aforementioned principle, a label is set for each pixel in the form of $\alpha/\beta$. In the label, $\alpha=\left\{0,1\right\}$ reflects the effective similarity, which equals 1  only when the two pixels at the same position of the two images are same and they are both important; $\beta=\left\{0,1\right\}$ reflects the amount of important information, which equals 1 when the pixel is important and equals 0 when the pixel is unimportant.

\begin{figure}[!t]
    \centering
    \includegraphics[width=2.1in]{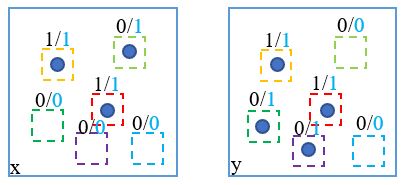}
\caption{Schematic diagram of the binary case. In image \textbf{\textit{x}} and image \textbf{\textit{y}}, the circles indicate important pixels, while the white parts around them are unimportant. Each label $\alpha/\beta$ above the dotted square indicates the state of the pixel.}
\end{figure}

\begin{figure}[!t]
    \centering
    \includegraphics[width=2.1in]{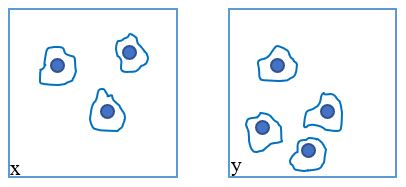}
\caption{Schematic diagram of the varying intensity case. In image \textbf{\textit{x}} and image \textbf{\textit{y}}, the circles indicate important objects, whose boundaries are the curves, while the white parts outside the curves are unimportant. Inside each object, the intensity of each pixel can be different.}
\end{figure}

Therefore, the similarity index for the two images in the binary case is given by Equation (\ref{binary}).
\begin{equation}
\label{binary}
    {LISI}_{binary}(x,y)=\frac{\sum\alpha}{\max\left( {\sum\beta_{x}},{\sum\beta_{y}} \right)}
\end{equation}

Generalising the binary case to the case of varying intensity, the schematic diagram is shown in Fig. 4. In the varying intensity case, the importance of each pixel is determined by its intensity, where higher intensity pixels are more important. In the label $\alpha/\beta$ in the varying intensity case, the effective similarity $\alpha$ should be large when the two pixels at the same position of the two images are similar and both with high intensities. So, $\alpha$ should be expressed as Equation (\ref{alph}), where $C$ is to avoid instability when $\left| {x_{i} - y_{i}} \right|$ is close to 0.
\begin{equation}
\label{alph}
    \alpha = \frac{D\left| {x_{i} + y_{i}} \right|}{\left| {x_{i} - y_{i}} \right| + C}
\end{equation}

The amount of important information, indicated by $\beta$, in the varying intensity case should relate to the intensity of each pixel; therefore, $\beta_{x} = x_{i}$ and $\beta_{y} = y_{i}$.

Hence, in the varying intensity case, LISI ($LISI \in \left\lbrack 0,1 \right\rbrack$, which equals 1 when the input images are identical) is developed as Equation (\ref{LISI}). The numerator illustrates the effective similarity, and the denominator reflects the amount of important information. Certainly, the inputs need to be normalised, as described in Equation (\ref{normalise}), in the pre-processing prior to the calculation of LISI.

\begin{equation}
\label{LISI}
LISI\left( {x,y} \right) = D\frac{\sum\limits_{i=1}^{N}\frac{\left| {x_{i} + y_{i}} \right|}{\left| {x_{i} - y_{i}} \right| + C_{1}}}{\max\left( {\sum\limits_{i=1}^{N}x_{i}},{\sum\limits_{i=1}^{N}y_{i}} \right) + C_{2}}
\end{equation}

In Equation (\ref{LISI}), constant variables $C_{1}$ and $C_{2}$ are to avoid instability when the denominators are close to 0. To set the value of $LISI$ to be between 0 and 1, and equal to 1 when inputs are identical, the factor $D$ should be $D = \frac{C_{1}}{2}$ and very small. Herein, we set $C_{1} = C_{2} = 1 \times {10}^{-4}$, so $D = 5 \times {10}^{-5}$, which are sufficiently small when comparing with pixel values.

\begin{figure*}[!t]
    \centering
    \subfloat[]{\includegraphics[width=1.7in]{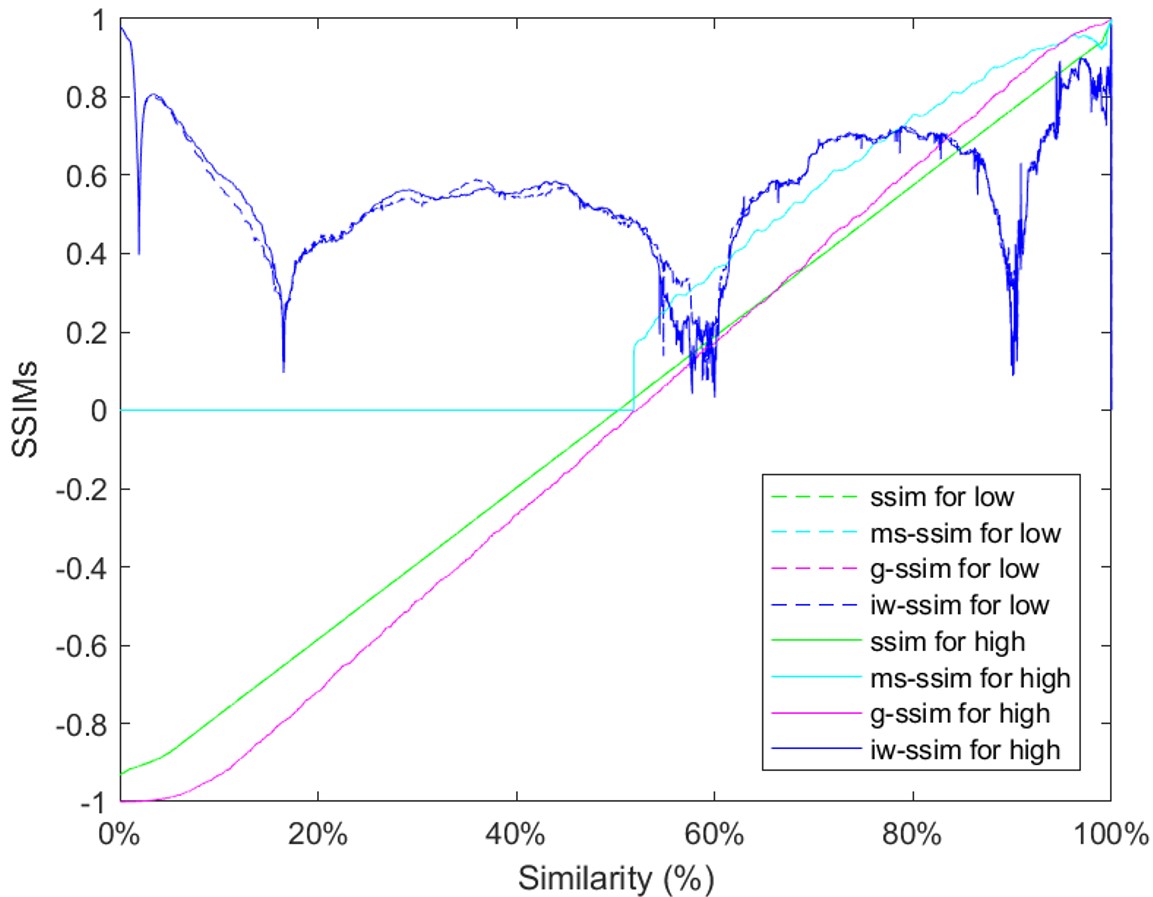}}
    \hfil
    \subfloat[]{\includegraphics[width=1.7in]{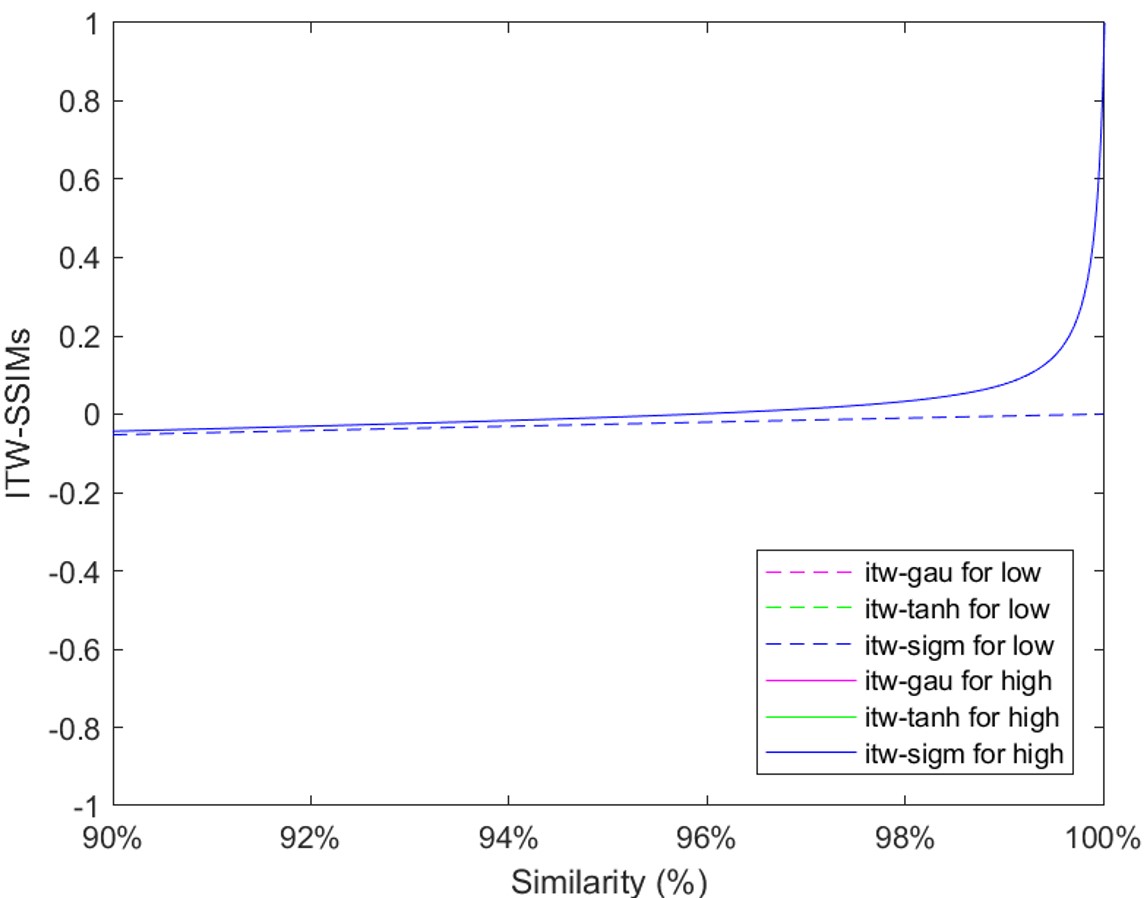}}
    \hfil
    \subfloat[]{\includegraphics[width=1.7in]{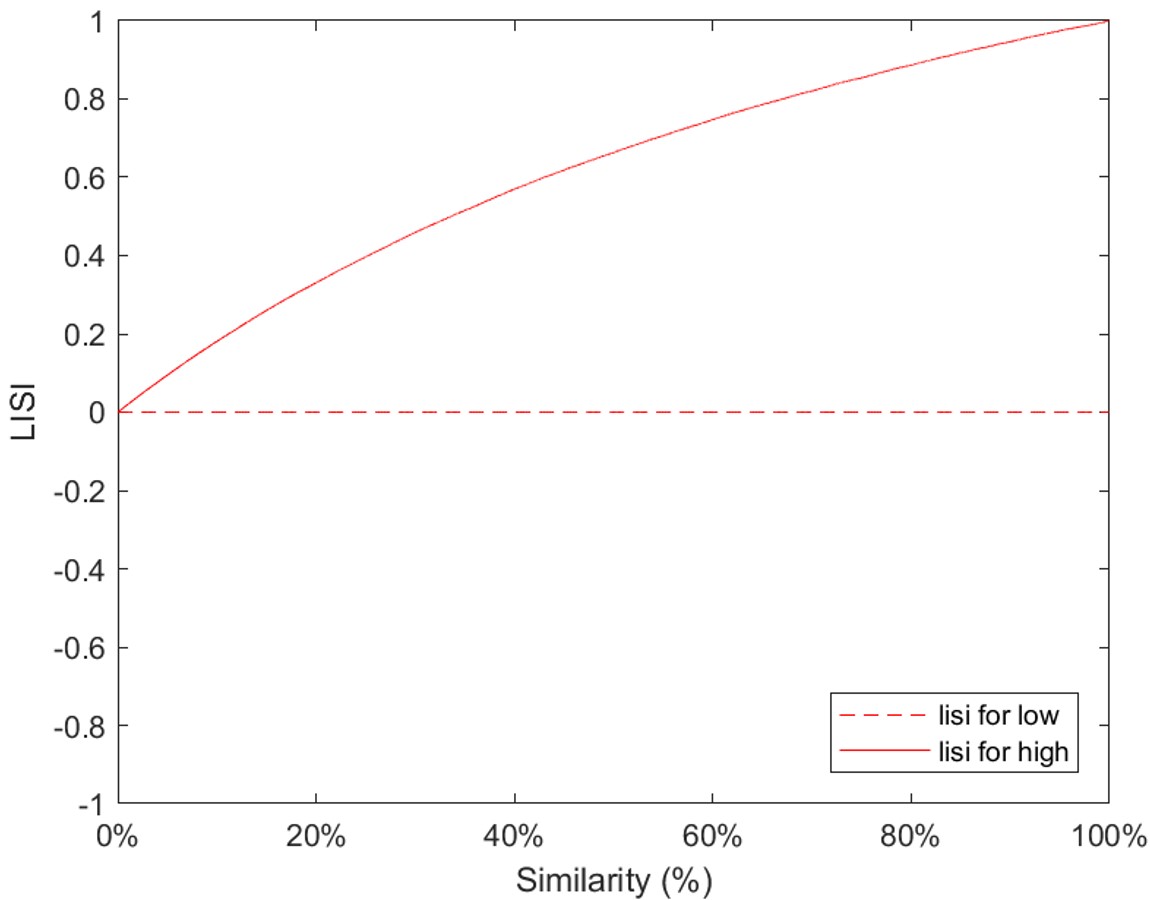}}
    \caption{Characteristic curves of (a) SSIMs, (b) ITW-SSIMs, and (c) LISI. In this figure, "for low (or high)" indicates IQA values for differences in low- (or high-) intensity parts. In (b), only the curves corresponding to the similarity higher than 90\% are shown, since ITW-SSIMs for input images with similarity lower than 90\% show similar sensitivity in both types of differences.}
\end{figure*}

\section{Experimental Results}

In order to evaluate the performance of ITW-SSIMs and LISI, natural images and field-specific images are used as inputs. Two auxiliary indexes are proposed in this section to assist the evaluation.

\subsection{Auxiliary Indexes}

\subsubsection{Sensitivity Index}

A sensitivity index $sensi$ is developed to evaluate the improvement in sensitivity of an IQA method (e.g., our new IQA methods) when comparing with a baseline (e.g., the original SSIM), which is expressed by Equation (\ref{sensi}). When the sensitivity of a new IQA method is higher than that of the original SSIM, $sensi$ is larger than 0; otherwise, it is smaller than 0. The larger its value is, the greater the improvement in its sensitivity.

\begin{equation}
\label{sensi}
    sensi \left( SSIM, IQA \right) = \frac{SSIM - IQA}{1 - SSIM}
\end{equation}

\subsubsection{Direction Index}

A direction index $direc$ is developed to characterise the direction of similarity in time-related processes, which is expressed by Equation (\ref{direc}). If $direc$ equals 1 (or -1), the intensity of the former input is larger (or smaller) than that of the latter input; if $direc$ equals 0, the two inputs are of similar intensity levels, generally.
\begin{equation}
\label{direc}
    direc\left( {x,y} \right) = \left\{ \begin{matrix}
{1,~\sum\limits_{i=1}^{N}\left( {x_{i} - y_{i}} \right) > 0} \\
{0,~\sum\limits_{i=1}^{N}\left( {x_{i} - y_{i}} \right) = 0} \\
{- 1,~\sum\limits_{i=1}^{N}\left( {x_{i} - y_{i}} \right) < 0} \\
\end{matrix} \right.
\end{equation}

\subsection{Characteristic Curves of IQA Methods}

To demonstrate the effectiveness of the proposed metrics, characteristic curves of IQA methods are shown in Fig. 5.

With the increase in similarity of two input images, the original, MS-, G-, and IW-SSIMs show similar sensitivity in differences in low- and high-intensity parts. On the contrary, in ITW-SSIMs with different weighting functions, ITW-SSIMs have higher sensitivity in differences in high-intensity parts especially when the two input images are very similar (with similarity greater than 90\%). Better still, LISI has higher sensitivity in differences in high-intensity parts across different similarities of the two input images. Thus, the new IQA indexes are more appropriate for intensity-sensitive cases.

\subsection{Natural Images}

To test the performance of the new IQA methods on natural images, reference images in the Tampere Image Database 2013 (TID2013, version 1.0) \cite{TID2,TID1,TID3} are used, as shown in Fig. S1 in the supplementary material.

Different types (uniform, Gaussian, and Rayleigh) of noise are added into each reference image and then the noisy images are compared with the corresponding reference image. The noisy images are divided into two groups, i.e., high-intensity and low-intensity noise images, where noises are only added into the highest or lowest $x\%$ ($x=35$ herein) intensity part of the image. Multiple noisy images with different noise positions are generated so that more reliable statistics can be calculated.

\begin{figure}[!t]
    \centering
    \subfloat[]{\includegraphics[width=1.6in]{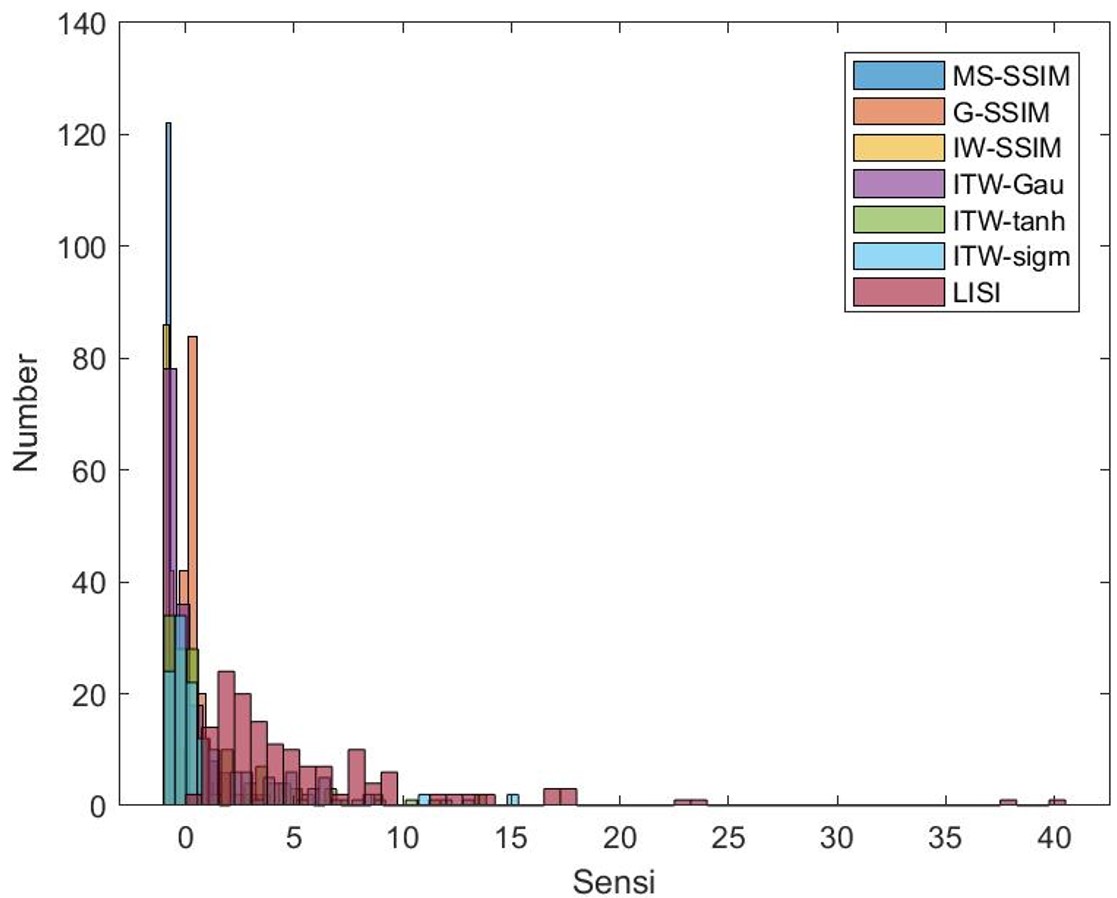}}
    \hfil
    \subfloat[]{\includegraphics[width=1.6in]{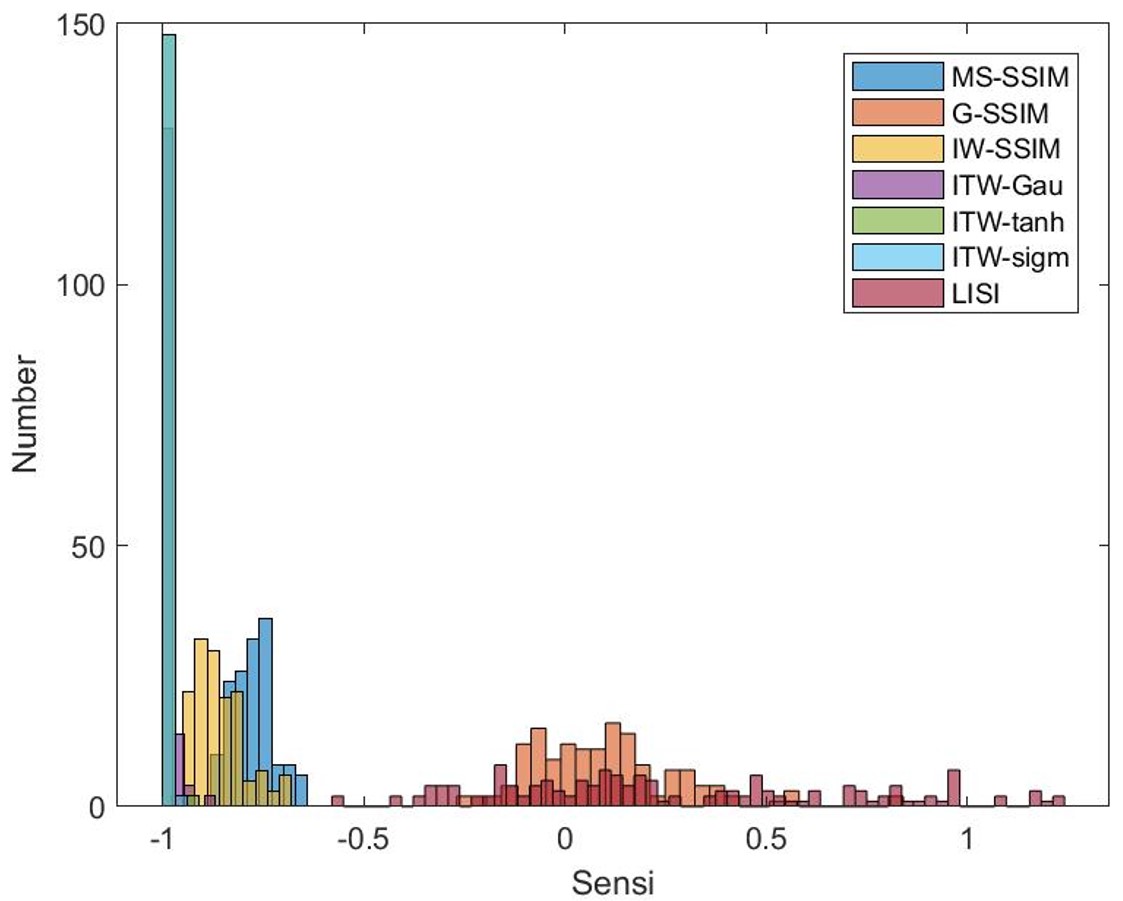}}
    \caption{Histograms of $sensi$ of inputs in (a) the high-intensity and (b) the low-intensity noise groups.}
\end{figure}

In the comparison of each pair of images (a noisy and its corresponding reference images), the SSIMs (original, MS, G, IW) and the new IQA methods (ITW-SSIMs and LISI) are used to assess the similarity. The improvement in sensitivity of the IQA methods comparing with SSIM is then evaluated by $sensi$. To visualise the distribution of $sensi$ for each IQA method performing on different inputs, histograms are generated for the two groups, as shown in Fig. 6. The statistical results of $sensi$ for the two groups are shown in Table S.I and S.II, where SD indicates standard deviation.

According to the results, in the high-intensity noise group, LISI is the most sensitive IQA method among the candidates, which performs much better than SSIMs. ITW-SSIM with tanh- and sigmoid-weighting are better than SSIMs to some extent, while in most cases the ITW-SSIM with Gaussian-weighting is worse than the original SSIM and G-SSIM but still better than MS- and IW-SSIMs. In the low-intensity noise group, the new IQA methods are less sensitive as expected since the high-intensity part of the images is of higher importance in new IQA methods while the differences in low-intensity part is less important. Nevertheless, LISI performs similarly to the original SSIM in most cases in the low-intensity noise group.

\subsection{Radio Astronomical Images}

Radio astronomy studies celestial phenomena by observing celestial sources in the radio wave band. In RA imaging, IQA methods are very useful. For example, the most pragmatic RA imaging algorithm, CLEAN \cite{CLEAN1,CLEAN2,CLEAN3}, is an iterative process used to extract point sources from the image obtained by inverse Fourier transforming the observed radio emissions. IQA methods can be used to quantify the change in the process. As a proof-of-concept experiment, the results of this usage of IQA methods in RA images are shown in this paper.

To illustrate the performance of ITW-SSIM and LISI on RA images, the restored images of GLEAM with different noise levels (15\%, 10\%, 5\%, 2\%) generated using the Hogbom CLEAN algorithm \cite{CLEAN1} in the Common Astronomy Software Applications (CASA) package \cite{CASA} are shown in Fig. S2 in the supplementary material. In these images, celestial sources only occupy a few parts of the images while the other parts of the images are noise-like.

To quantify the change in the iterative process, the results of different IQA methods applied to these images are shown in Table S.III, where three comparisons are made, i.e., 15\% vs 2\% noise-level (NL) images, 10\% vs 2\% NL images, and 5\% vs 2\% NL images. The data in Table S.III is the similarity (indicated by "Sim"). For the example images shown in Fig. 1, the original SSIM is 0.9707, whereas ITW-SSIM with Gaussian-, tanh-, and sigmoid-weighting are 0.9455, 0.9513, and 0.9635, respectively, and LISI is 0.0309. Following that, $sensi$ and $direc$ are calculated for each comparison, as shown in Table S.IV. As the iterative process proceeds, the direction index shows that the noise in the image is reduced. According to the results, LISI is the most sensitive similarity index to assess radio astronomical images, where the improvement of sensitivity becomes more obvious when the two input images are more similar. In contrast, ITW-SSIMs with the three weighting functions have similar performance and all better than SSIMs, where the improvement of sensitivity becomes more obvious when the two input images are more different. On the contrary, MS-, G-, and IW-SSIMs are not suitable for assessing radio astronomical images.

\subsection{Medical Images}

IQAs are useful in medical image assessment \cite{medi}. In medical images, intensity is often related to lesions. As many medical images show the low-information property, the new IQA methods can be helpful when dealing with such images. 

To test the performance of ITW-SSIM and LISI on medical images, a group of data (Subject ID: Lung\_Dx-A0029) from Lung-PET-CT-Dx dataset \cite{lung,lung1} is used, which comprises computed tomography (CT) and Positron emission tomography (PET-) CT images of lung cancer subjects. Pursuant to a tissue histopathological diagnosis, the patient in this group of images was diagnosed with Adenocarcinoma. The images are taken at 4 different time points. After registration, the images are shown as Fig. S3 in the supplementary material.

To quantify the change with time, the results of different IQA methods on these images are shown in Table S.V, where four comparisons are made. In the first three comparisons, images at adjacent time points are compared; while in the fourth comparison, the earliest image and the latest image are compared. Following that, $sensi$ and $direc$ are calculated for each comparison, as shown in Table S.VI. The change happening in the medical images is visualised in Fig. 7. The trend of the polyline is defined by the direction index. It goes down when the intensity increases while it goes up when the intensity decreases. The value of the difference is complementary to the similarity indicated by the IQA indexes. Crops are made, as shown in Fig. 8 (a), to characterise the change in each part of the patient's lung whose results are shown in Fig. 8 (b) to (d).

\begin{figure}[!t]
    \centering
    \includegraphics[width=2.9in]{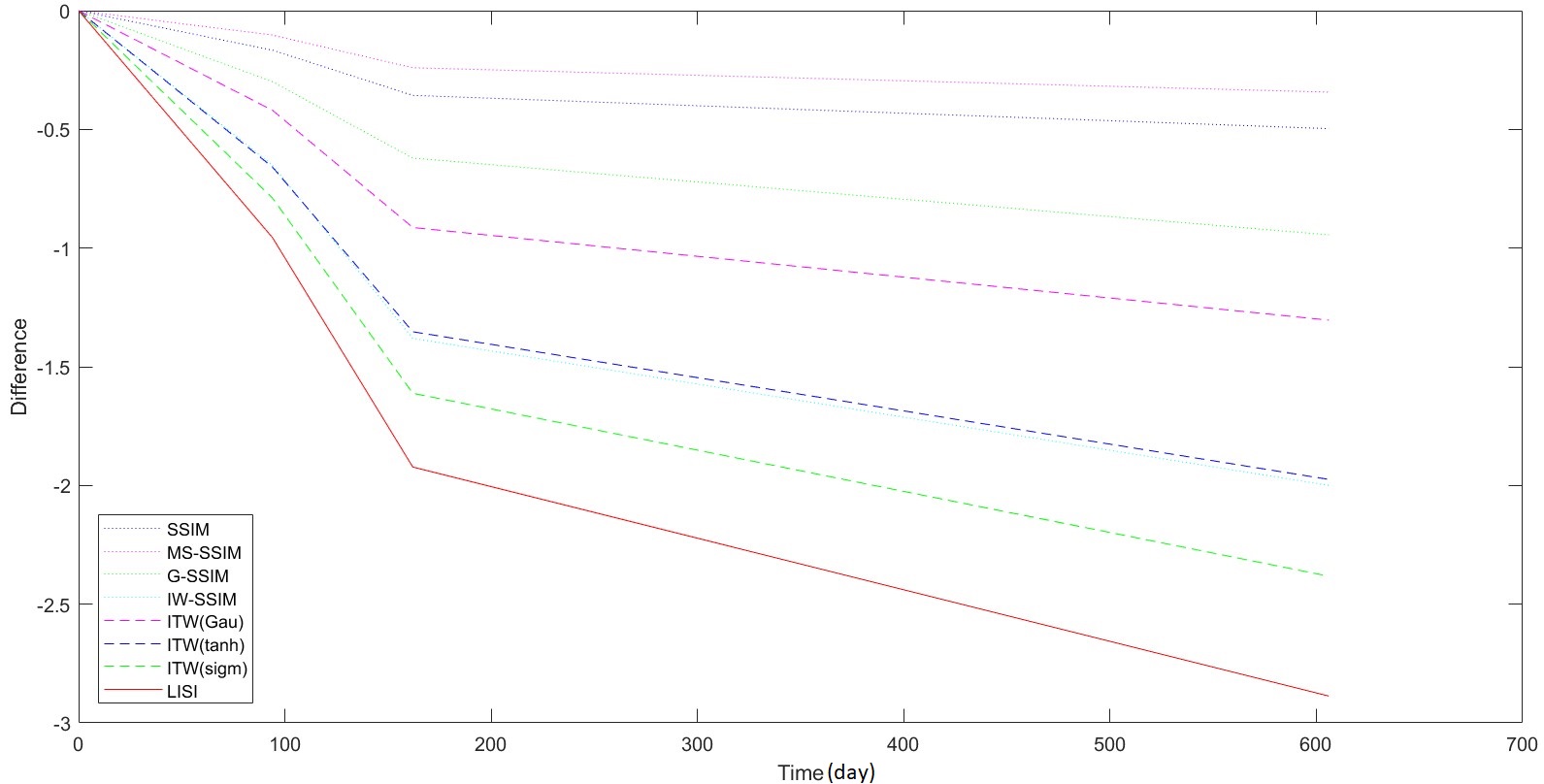}
    \caption{Change of medical images.}
\end{figure}

\begin{figure}[!t]
    \centering
    \subfloat[]{
        \includegraphics[width=1.2in]{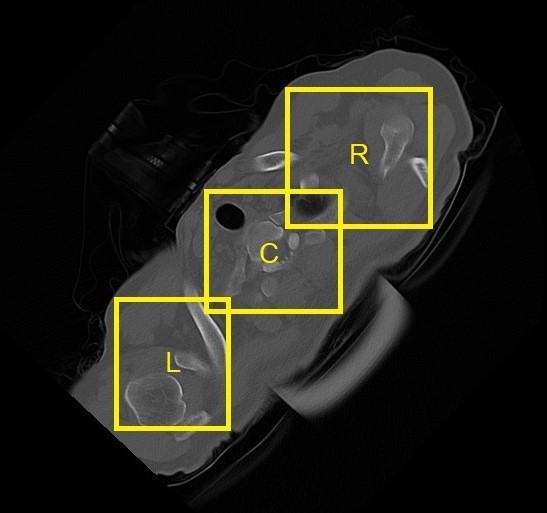}}
    \hfil
    \subfloat[]{
        \includegraphics[width=1.45in]{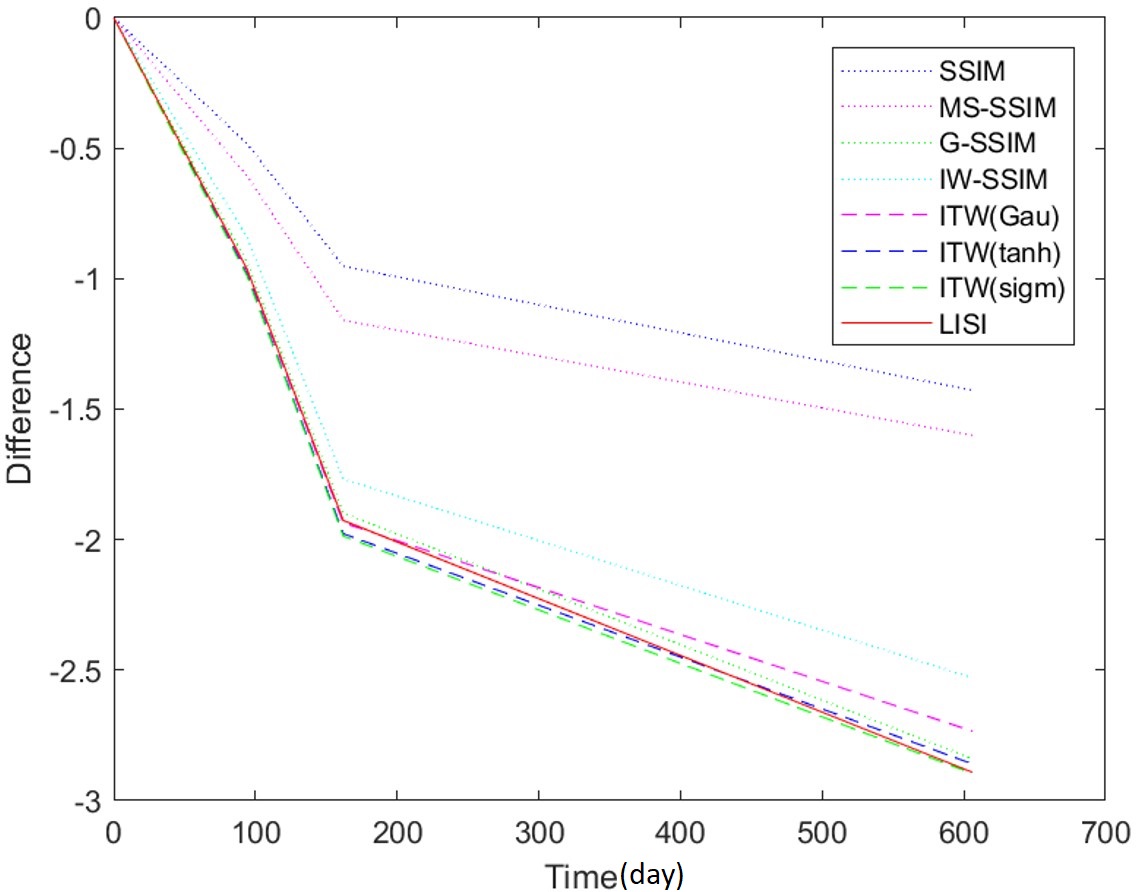}}
    \hfil
    \subfloat[]{
        \includegraphics[width=1.45in]{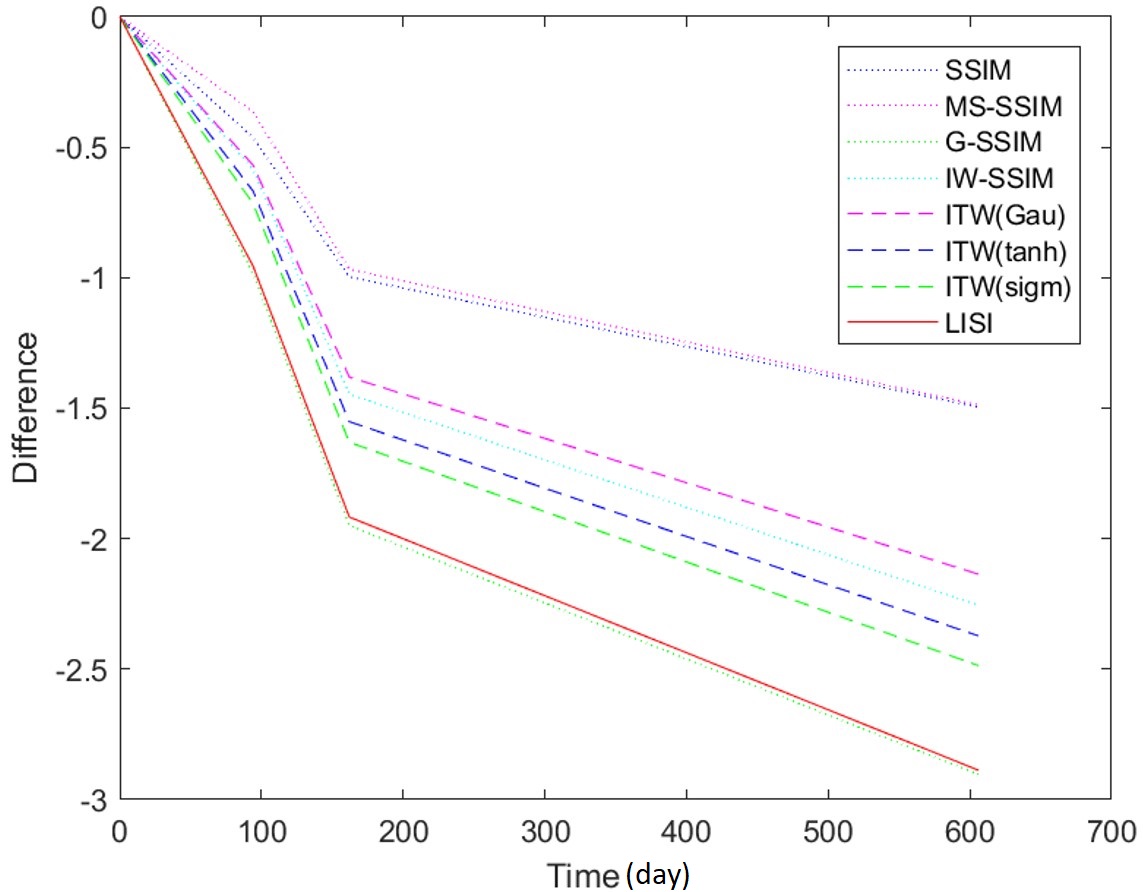}}
    \hfil
    \centering
    \subfloat[]{
        \includegraphics[width=1.45in]{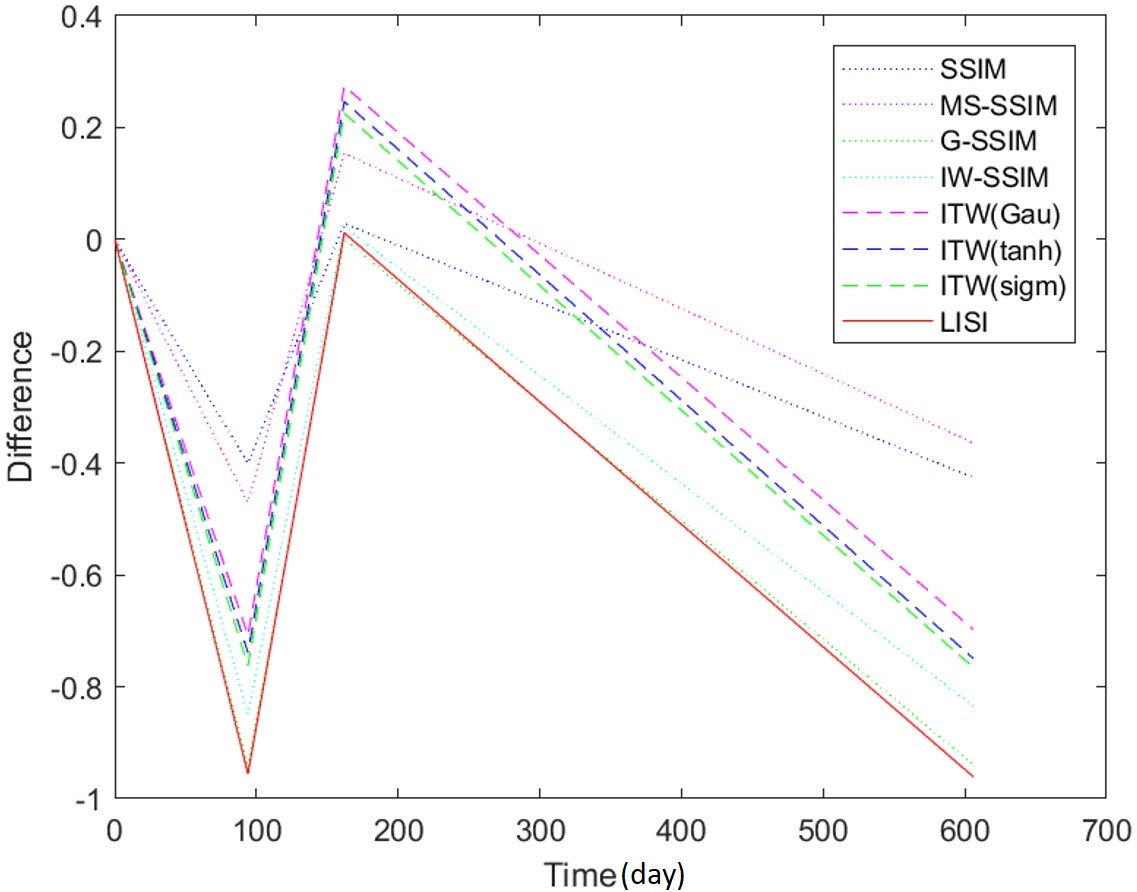}}
    \caption{(a) Cropped medical images, where "L", "C", and "R" indicate the left, central, and right crops of the patient's lung; and, change of difference in (b) L, (c) C, and (d) R.}
\end{figure}

According to the results, generally, LISI is the most sensitive similarity index to assess this group of medical images, and ITW-SSIM with sigmoid weighting also works well. The performance of the other two ITW-SSIMs are also better than that of SSIM, while IW-SSIM performs fine as well. The direction index shows that the total intensity of the image increases with the time. In the left crop, the new IQA methods and G-SSIM have similar performance and are much better than the performance of the original SSIM and MS-SSIM; nevertheless, all of them show the same trend of the intensity changing in the left crop. In the central crop, it shows that LISI and G-SSIM are two best choices for the assessment. Interestingly, in the right crop, the intensity fluctuates with the time. From Fig. 8 (d), LISI is more sensitive when the intensity increases, while ITW-SSIMs are more sensitive when the intensity decreases.

\subsection{Remote Sensing Images}

In remote sensing (RS), it is useful to use IQA methods. For example, by assessing images of dryness, early warning of fire can be improved. The low-information property can be ensured by inversion of intensity in these images.

To demonstrate the performance of ITW-SSIM and LISI on RS images, a group of data from a weekly geospatial dataset QuickDRI \cite{RS} is used, which indicates short-term dryness across the United States. California has suffered from many wildfire incidents in recent years. To assess and analyse the dryness of California, we divide the images of California into 12 regions according to their geographical locations, as shown in Fig. 9 (a). In the images, darker parts are drier, while brighter parts are wetter. After registration, the dryness images of California collected from 02 Nov 2020 to 31 Oct 2021 are shown in Fig. S4 in the supplementary material. The changes happening in the 12 regions are visualised in Fig. S5 in the supplementary material. The trends of the polylines, where different IQA methods are shown in different marks, are defined by the direction index. The line goes down when the region becomes wetter while goes up when the region becomes drier. The difference is obtained by comparing the images from adjacent weeks, where the value of the difference is complementary to the similarity indicated by the IQA indexes.

\begin{figure}[!t]
    \centering
    \includegraphics[width=2.8in]{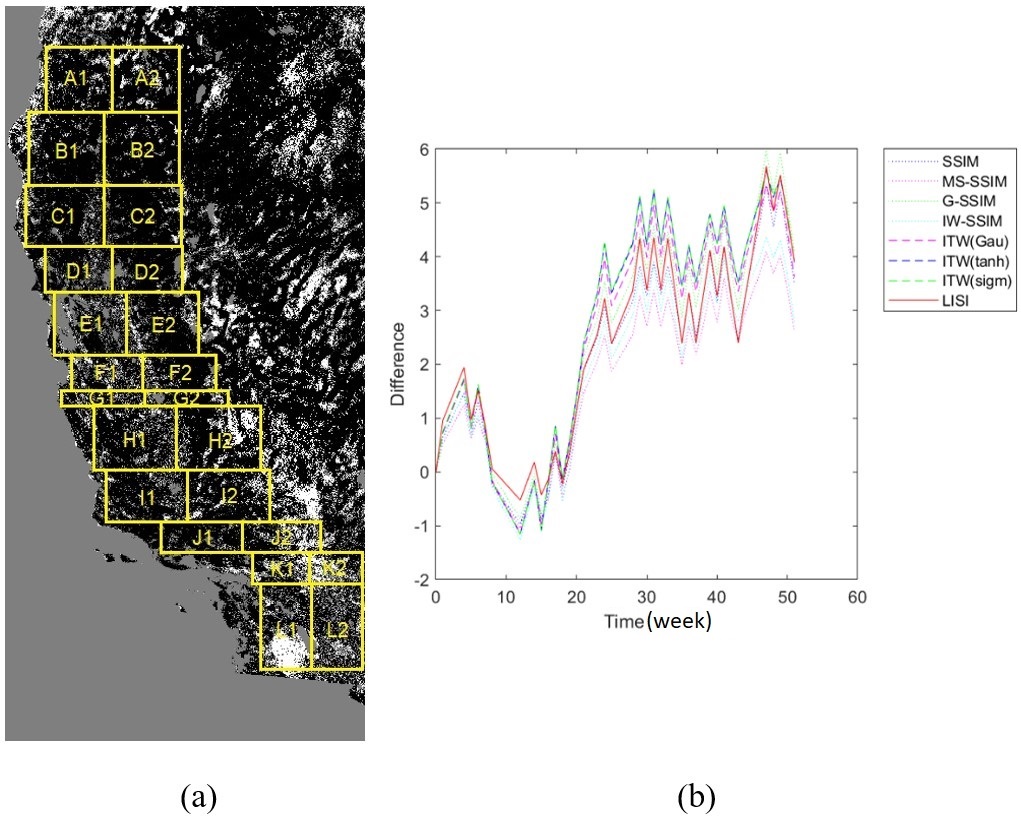}
    \caption{(a) Regions of California; (b) Change of dryness in region B2.}
\end{figure}

By analysing the results, some early warning of fire can be obtained. The tendency of drought may warn people of fire risk. For example, in Region B2, Laura 2 fire started in Lassen County on 17 Nov 2020 (contained on 24 Nov 2020), which destroyed 48 structures, damaged 4 structures, and burnt 2,800 acres \cite{lassen1,lassen2,lassen3}; Beckwourth Complex fires started in Plumas and Lassen Counties on 3 Jul 2021 (contained on 22 Sep 2021), which destroyed 148 structures, damages 23 structures, and burnt 105,670 acres \cite{pl1,pl2}; Dixie Fire started on 13 Jul 2021 (contained on 25 Oct 2021), which involved Lassen (B2), Plumas (B2), Butte (B2), Tehama (B1), and Shasta (B1) Counties, destroyed 1,329 structures, damaged 95 structures, and burnt 963,309 acres \cite{pl1,dixie1,dixie2,dixie3,dixie4}. The dryness of region B2 is illustrated in Fig. S5 (d) (restated in Fig. 9 (b)). Laura 2 fire has shown a tendency of drought since the week 09-15 Nov 2020 by LISI, and the tendencies indicated by ITW-SSIMs are also more obvious than SSIM. Beckwourth Complex fires have shown a tendency of drought since the week 14-20 Jun 2021 by ITW-SSIM with sigmoid- or tanh-weighting, which was at least one week earlier than SSIM. Also, Dixie Fire has a more obvious tendency in ITW-SSIM with sigmoid- or tanh-weighting.

According to the results, including, but not limited to, the aforementioned examples, for RS images, ITW-SSIM with sigmoid- and tanh-weighting perform the best, which may provide valuable early warning for disasters such as wildfires.

\section{Conclusion}

New field-specific IQA methods, i.e., ITW-SSIM and LISI, are developed in this paper, which are more sensitive to intensity than SSIMs to assess the similarity between two low-information images. Two auxiliary indexes, sensitivity index and direction index, are proposed to evaluate the improvement of sensitivity and characterise the direction of similarity in assessing sequential images that describe the change of a process. ITW-SSIMs and LISI are applied to natural images and field-specific images in the experiments. As comparison, the original, MS-, G-, IW-SSIMs are applied to those data as well. According to the results, it can be concluded that our IQA methods have better performance than SSIMs generally, and ITW-SSIMs and LISI are suitable for different applications. ITW-SSIM and LISI can be applied in various fields, such as radio astronomical imaging, medical imaging, remote sensing, and so on.

\section*{Acknowledgment}

The authors would like to thank Dr. Fred Dulwich, Dr. Ben Mort, and Dr. Karel Adamek for helpful discussions regarding radio astronomy.

\bibliographystyle{IEEEtran}
\bibliography{Bibliography.bib}

\end{document}